\documentclass[%
 reprint,
 superscriptaddress,
 amsmath,amssymb,
 aps,
 prl,
]{revtex4-2}

\usepackage{xcolor}

\newcommand{\etal}{\textit{et al}}

\usepackage{graphicx}
\usepackage{dcolumn}
\usepackage{bm}

\usepackage{booktabs} 		
\usepackage{etoolbox}

\DeclareUnicodeCharacter{0301}{}  

\begin{document}

\preprint{APS/123-QED}

\title{
  Prediction of NMR, X-ray and Mössbauer experimental results for amorphous Li-Si alloys using a novel DFTB model}

\author{F. Fernandez}
\author{M. Otero}
\affiliation{
 Universidad Nacional de Córdoba, Facultad de Matemática, Astronomı́a, Fı́sica y Computación, Córdoba (X5000HUA), Argentina
}%
\affiliation{
 Consejo Nacional de Investigaciones Cientı́ficas y Técnicas (CONICET), Instituto de Fı́sica Enrique Gaviola, Córdoba (X5000HUA), Argentina
}%

\author{M. B. Oviedo}%
\affiliation{%
 Universidad Nacional de Córdoba, Facultad de Ciencias Quı́micas,
Departamento de Quı́mica Teórica y Computacional, Córdoba (X5000HUA),
Argentina
}%
\affiliation{%
 Consejo Nacional de Investigaciones Cientı́ficas y Técnicas (CONICET),
Instituto de Fisicoquı́mica de Córdoba (INFIQC), Córdoba (X5000HUA), Argentina
}%

\author{D. E. Barraco}
\affiliation{
 Universidad Nacional de Córdoba, Facultad de Matemática, Astronomı́a, Fı́sica y Computación, Córdoba (X5000HUA), Argentina
}%
\affiliation{
 Consejo Nacional de Investigaciones Cientı́ficas y Técnicas (CONICET), Instituto de Fı́sica Enrique Gaviola, Córdoba (X5000HUA), Argentina
}%

\author{S. A. Paz}%
\email{apaz@unc.edu.ar}
\author{E. P. M. Leiva}%
\affiliation{%
 Universidad Nacional de Córdoba, Facultad de Ciencias Quı́micas,
Departamento de Quı́mica Teórica y Computacional, Córdoba (X5000HUA),
Argentina
}%
\affiliation{%
 Consejo Nacional de Investigaciones Cientı́ficas y Técnicas (CONICET),
Instituto de Fisicoquı́mica de Córdoba (INFIQC), Córdoba (X5000HUA), Argentina
}%



\date{\today}

\begin{abstract}


Silicon anodes hold great promise for next-generation Li-ion batteries. The main obstacle to exploiting their high performance is the challenge of linking experimental observations to atomic structures due to the amorphous nature of Li-Si alloys. We unveil the atomistic-scale structures of amorphous Li-Si using our recently developed density functional tight-binding model. Our claim is supported by the successful reproduction of experimental X-ray pair distribution functions, NMR and Mössbauer spectra using simple nearest neighbors models. The predicted structures are publicly available.

\end{abstract}

\maketitle



Global warming caused by fossil fuel burning appears as the biggest 
environmental problem facing us in this century. One of its major 
contributions is due to the use of internal combustion vehicles, so a 
transition to electric vehicles maintaining autonomy and charging time 
are of vital importance along with habit changes. 
One of the goals set by the Intergovernmental Panel on Climate Change 
(IPCC) is to limit the average temperature increase to 1.5$^{\circ}$C. 
This requires changes in technology and human behavior, one of the 
most important of which is that by 2050, 80\% of energy must be supplied 
by renewable sources~\cite{harvey2018}. Next-generation lithium batteries 
are the most promising option to meet the intermittency of renewable energies 
and power electric vehicles, but they require an increase in 
their capacity. In this sense, silicon anodes are presented as the best 
candidate because of the high theoretical capacity of 3579 mAhg$^{-1}$,
which is ten times higher than the current graphite anodes, besides being
a cheap, abundant and environmentally friendly material. However, it 
presents large volumetric changes, of the order of 300\%, 
which lead to structural degradation and a consecutive capacity lost in 
successive charge/discharge cycles~\cite{obrovac2004}.

It has been shown~\cite{liu2019} that knowledge at the atomic level of
batteries' active materials allow designing strategies that mitigate their
limitations and greatly improve performance. This has inspired the scientific
community to apply numerous microscopic and spectroscopic characterization
techniques. The intrinsic behavior of Si anodes leads to the formation of
amorphous Li-Si alloys during charge/discharge that make the short-range
structure especially relevant. It has been stated that the
crystal-amorphous phase transition occurring for this system represents the main obstacle to improving its electrochemical performance, mainly because it hinders the
attempts to link the atomic structures with the experimental
observations~\cite{key2011}. Although there are experiments related to local
structures such as nuclear magnetic resonance (NMR), Mössbauer spectroscopy (MB),
X-ray pair distribution function (PDF), among others, their interpretation is
evasive without a precise theoretical model capable of predicting the microscopic
structure of the system and correlate it with the experimental observables. As
an example, two different states of Li-ions are always observed, and an
intermediate transition that has not yet been elucidated, in
voltammograms~\cite{pan2019}, in diffusion coefficients~\cite{ding2009} and in
NMR experiments~\cite{key2009}.

In previous work, we parametrized a DFTB potential that exhibits a remarkable
precision in the prediction of the formation energies for several Li$_x$Si
crystalline and amorphous structures, in a wide range of
compositions~\cite{oviedo2022}. Furthermore, using this potential, the simple
annealing of crystalline silicon (c-Si) resulted in an amorphous structure
(a-Si) with a RDF that is in perfect agreement with experimental
results~\cite{laaziri1999}. In light of these results, we show here that the
amorphous structures predicted by our model can be used to understand
NMR~\cite{key2009, key2011, koster2011, ogata2013}, MB~\cite{li2009} and X-ray
PDF~\cite{wang2020} experimental measurements using a simple nearest-neighbors
model.
 
We started by lithiating the a-Si structure obtained in our previous
work~\cite{oviedo2022}, following a procedure similar to that proposed by
Chevrier and Dahn~\cite{chevrier2009}: (i) Add a Li atom at the center of the
largest spherical void, increase the volume and scale coordinates by a factor
to obtain the experimental expansion in the system; (ii) Perform an NPT
molecular dynamics equilibration for 10ps using the Berendsen Thermostat and
Barostat~\cite{berendsen1984} available in the DFTB+ code~\cite{dftb+};
repeat these two steps until the desired number of Li atoms is reached.
Step~(ii) represents a slight modification that improves the fixed-volume
coordinate optimization procedure performed at reference~\cite{chevrier2009}.
Following this procedure, we obtain amorphous structures for a wide range of
Li$_x$Si concentrations, from the initial 64 Si atoms ($x=0$) to a total of 304
for the fully lithiated structure ($x=3.75$). These obtained structures are available in a public repository \cite{}.

First, we characterize the amorphous structures by computing the partial radial
distribution functions (RDF), which describe the probability of finding an atom
in a shell at a distance $r$ from a reference atom. As can be seen in Figure~S1
of Supplemental Material, the nature of the obtained distributions 
is typical for amorphous structures, having a definite first peak at short $r$
and decreasing following ones for increasing $r$. 
The most interesting behavior to analyze is found in the Si-Si RDF. As lithium
concentration increases there is clearly a decrease in the first neighbor peak
together with a shift in the second neighbor peak towards larger distances (3.8
\AA\ $\rightarrow 4.7 $\AA). This tells us that the Si--Si bonds are being
broken during the lithiation and isolated Si appears.
It is worth mentioning that the DFTB model used in this work allows to capture
these fine chemical features at a low computational cost, showing a great
detail in the representation of the electronic structure of many atoms. 

\begin{figure}
    \centering
    \includegraphics[width=\columnwidth]{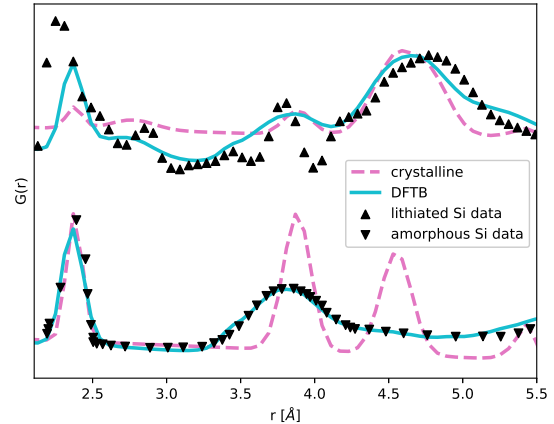}
    \caption{\label{fig:gofrs} Pair distribution functions $G(r)$ for amorphous
      and full lithiathed Si contrasted with experimental data. The triangles
        pointing down correspond to Laaziri's a-Si
        experiment~\cite{laaziri1999}, while those pointing up correspond to
				Key \etal.'s experiment~\cite{key2011}.}
\end{figure}

Combining the previous partial RDFs allowed us to calculate the total $G(r)$ for each
structure, as detailed in the Supplemental Material. The $G(r)$ can be
directly compared with the PDF obtained from X-ray measurements~\cite{billinge2019}. 
The standing triangles of
Figure~\ref{fig:gofrs} show the X-ray PDFs for the two extreme cases of unlithiated a-Si (bottom) and full lithiated silicon anode (top), as measured by Laaziri
\etal.~\cite{laaziri1999} and Key~\etal.~\cite{key2011}, respectively. For
comparison, we include in the same figure the $G(r)$ computed using the
following procedures. For a-Si, we directly take the $G(r)$ as an
average from our a-Si modeled structures, resulting in an excellent agreement
with the experimental measurement, as can be seen at the bottom of the figure. For the case
of the fully lithiated silicon anode, the experimental sample is expected to be mainly composed of
amorphous Li$_{15}$Si$_4$, although a crystalline contribution to the X-ray PDF
may also be expected. In addition, another contribution from pure Si is also
possible, mainly due to an incomplete lithiation resulting from
various experimental factors such as bad connectivity, kinetically limited events, or even a
possible Li$_{15}$Si$_4$ decomposition~\cite{key2009, key2011}. Therefore, we
have fitted the experimental data shown in figure~\ref{fig:gofrs} using a linear
combination of the $G(r)$ obtained from the a-Li$_{15}$Si$_4$ and
a-Si modeled structures, but also including the crystalline ones (see
Supplemental Material). Strikingly, the resulting weights for the crystal phases
are quite small, representing only a 9.78\% for c-Li$_{15}$Si$_4$ and exactly 0\% for
c-Si. This highlights the importance of counting with amorphous atomic
structures to understand the experimental measurements. On the contrary, a very poor fit
is obtained if only crystal structures are used, as can be seen at the top of
figure \ref{fig:gofrs} (pink curve).  



Direct measurement of local atomic structure is challenging, however, some
spectroscopic techniques allow us to infer the local structure due to a high
dependence of the measured property on the atomic local environment. This is
the case for NMR and MB spectroscopy. In previous work, Key
\etal.~\cite{key2009} prepared different Li-Si structures and measured their
$^7$Li NMR chemical shift spectra. The NMR signal is characterized by a Voigt
peak ($V$), which corresponds to the combination of a Lorentzian that is
intrinsic to the NMR phenomena and a Gaussian due to the
detector~\cite{higinbotham2001}. Each particular nucleus introduces a shift on
this signal, the chemical shift, that depends on the electromagnetic shielding
caused by the local environment, and therefore the local structure.  The total
spectra obtained for each sample are the sum of the signal contributions coming
from all the $^7$Li in the structure. In this spectra,
Key~\etal.~\cite{key2009} assigned a peak at 18 ppm to the chemical shift of a
$^7$Li atom which is near a bonded Si atom. On the other hand, a peak at 6 ppm
was attributed to $^7$Li near an isolated Si atom. However, it is not clear how
to interpret the occurrence of peaks between 6 and 18 ppm, which are in fact
observed in the NMR measurements.  We propose here a simple nearest-neighbor
model to emulate and interpret these features in the NMR spectra. We define a
peak position in the spectra including an average of the contributions stemming from all Si
atoms sitting in the first coordination
shell of a given Li atom as follows:
\begin{equation}
    \delta_{Li} = \frac{1}{N_{Si}} \sum_{Si \in \text{NN}} \delta_{\text{Key}},
\end{equation}
where $\delta_{Li}$ is the chemical shift contribution of the Li atom 
in question to the global spectra, the sum is considered over all first 
nearest neighbours (NN) Si atoms ($N_{Si}$) and $\delta_{\text{Key}}$ is 
the shift proposed by Key \etal.~\cite{key2009}, as discussed above. Then, the total chemical 
shift spectrum has an intensity ($I$) that results from a sum of Voigt peaks ($V$)  
over all the Li atoms in the structure $S$,
\begin{equation}
    I = \sum_{Li \in S} V(\text{ppm}, \delta_{Li}, \sigma, \gamma),
\end{equation}
where ppm is the chemical shift , $\sigma$ and $\gamma$ are the standard 
deviations and the half-width at half-maximum of gaussian and lorentzian 
components, respectively. Finally, an average was taken over all snapshots
of our molecular dynamics trajectories.

\begin{figure}
    \centering
    \includegraphics[width=\columnwidth]{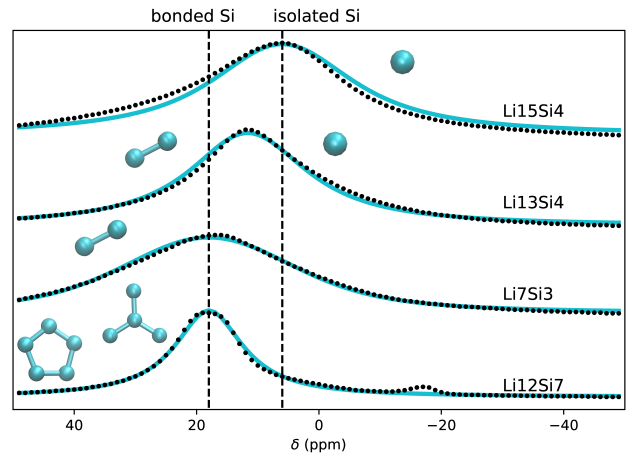}
    \caption{\label{fig:crystals} $^7$Li chemical shift spectra for 
    crystalline alloys. The dots correspond to the measurements of Key \etal.\cite{key2009}
 and the solid lines to the present nearest-neighbors model. 
    Vertical lines indicate the contributions of bonded 
    and isolated Si.}
\end{figure}

As a first test of this model, we present the $^7$Li chemical shift spectra for the crystalline structures in Figure \ref{fig:crystals}, as calculated from the present assumptions. The 
atomic configurations were obtained from the Materials Project~\cite{matproj}
and the peak width was fitted to the accuracy of the Key \etal. 
experiment~\cite{key2009} once the nearest-neighbors model determined the 
center of the peak. The model presented here corresponds to the 
continuous lines, while Key's measurements correspond to the points.
For all four alloys, the center of the peak is reproduced 
with high accuracy. In the Li$_{12}$Si$_{7}$ 
and Li$_{7}$Si$_{3}$ alloys, the Si atoms are all bonded. The first forms 
pentagons of 5 atoms or stars of 4, while in the second one, all the 
bonds are dumbbells. This shows that in this nearest-neighbors model, all the Li 
atoms contributions are centered at 18 ppm. In the fully lithiated alloy Li$_{15}$Si$_{4}$
there are only isolated Si atoms, so all the contributions are centered at
6 ppm. Meanwhile, for Li$_{13}$Si$_{4}$, isolated Si coexists with dumbbells,
making intermediate contributions to the spectra, consistent with experiment.

\begin{figure}
    \centering
    \includegraphics[width=\columnwidth]{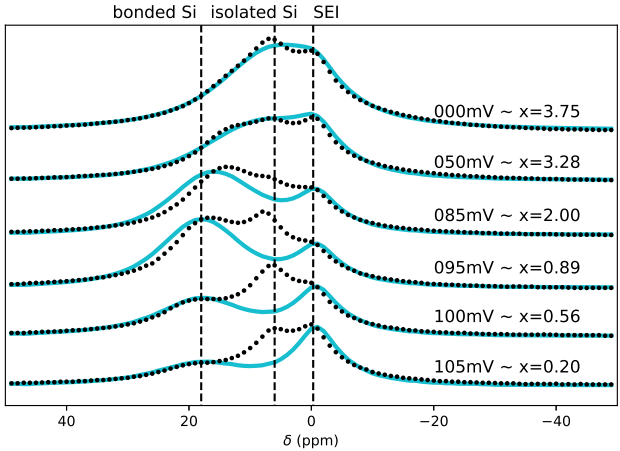}
    \caption{\label{fig:amorphous} $^7$Li chemical shift spectra for 
    amorphous structures. The dots correspond to the measurements of Key \etal.\cite{key2009}.
    The results of the present model are represented with solid lines and a SEI
    contribution is added for comparison. The vertical lines indicate the 
    contributions of SEI and bonded or isolated Si atoms.}
\end{figure}

Having obtained these promising results for the crystalline structures, we now
focus on amorphous structures, which are usually found in electrochemical
experiments. 
We calculate the $^7$Li chemical shift spectra for different amorphous Li-Si alloys with the proposed nearest-neighbors model using the structures obtained from our molecular dynamics simulation.
We considered
those structures with $x$ values in Li$_x$Si that can be related to the
voltages used by Key \etal.~\cite{key2009} to collect the NMR spectra. To
facilitate the comparison with our model we have included a peak at -0.3 ppm,
which represents the SEI contribution, as suggested by the cited authors.  The
results are shown in Figure~\ref{fig:amorphous} which, apart from some
differences explained below, shows an excellent agreement between the
computational and experimental results. This is particularly true in the case
of the peak at 18 ppm, where the modeling is able to mimic the shift of the
peak to 6 ppm at high concentration. This change is the clearest evidence to
support the current atomistic view of the system. At the beginning of the
lithiation, the configurations contain mainly bonded Si atoms. Then there is an
intermediate coexistence between bonded and isolated Si atoms. Finally,
isolated Si atoms prevail at high lithium concentrations. 

There is a better agreement between the model and experiment at high Li
concentrations. At low Li concentrations, the experiment shows an extra
contribution to the spectrum at 6 ppm, which is attributed to isolated Si atoms
due to an inhomogeneous electrochemical lithiation~\cite{key2009}, not present
in the simulated lithiated structures. These simulations correspond to
structures in thermodynamic equilibrium and do not take into account the
processes that may occur in an inhomogeneous lithiation of the electrode, where
highly lithiated regions exist at low overall lithium concentrations.

\begin{figure}
    \centering
    \includegraphics[width=\columnwidth]{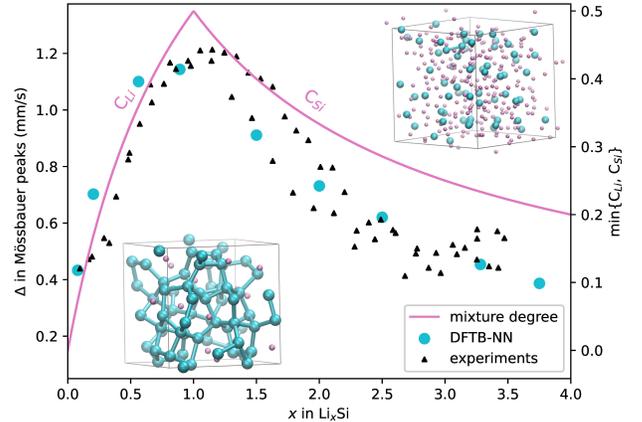}
		\caption{\label{fig:mossbauer} Shift between the two peaks in the Mössbauer
effect spectra. The triangles pointing up correspond to two measurements from
Li \etal.~\cite{li2009} (left axis), the solid line is the  prediction of equation \ref{eq:mossbauer}, using average concentrations of Li and Si atoms. The cyan circles are the prediction given by equation \ref{eq:mossbauer}, with  $C_{Li}$, $C_{Si}$ calculated from the nearest-neighbors concentration (right axis).} 
\end{figure}

A different perspective on the study of local atomic structure
was provided by Li \etal.~\cite{li2009}. They used gamma rays to measure the $^{119}$Sn
Mössbauer spectra (MB) of amorphous Li$_x$Si$_{1-y}$Sn$_y$, for $0<x<3.5$ and
small values of $y$~\cite{li2009}. Given the low concentration of tin, the
authors assume that Sn atoms occupy the same sites of Si atoms in this
material, making it equivalent to the result of an amorphous Si
lithiation~\cite{hatchard2005}.  The MB signal consists of two peaks that almost completely overlap at the extreme cases of low or high lithium
concentration, but are clearly separated for the intermediate cases. The
separation distance, $\Delta$, is expected to be sensitive to the local
environment of Sn atoms. $\Delta$ reaches a maximum value around 1.2 mm$/$s at $x
\sim 1$ and decreases for lower and higher concentrations, see triangles in Figure \ref{fig:mossbauer}. The authors suggest that the maximum value of $\Delta$ is obtained when Sn atoms
(and by analogy Si atoms) are surrounded by an equimolar mixture of Si and
Li, and then decreases when some of the two atom types predominate. 
In order to make a quantitative statement in terms of this idea, let us define 
the concentration of Li atoms, say $C_{Li}$ and the concentration of Si atoms, say $C_{Si}$ in terms of the number of Li atoms in the formula Li$_x$Si:
\begin{equation}\label{eq:C}
C_{Li} = \frac{x}{1+x} \hspace{0.8cm}; \hspace{0.8cm}  C_{Si} = 1 - C_{Li} 
\end{equation}
Now, since we find from the experimental results in Figure \ref{fig:mossbauer} that $\Delta$ tends to a constant value at small and large $x$ values, we propose for $\Delta$ the following ansatz:
\begin{equation}\label{eq:mossbauer}
    \Delta = a\min\left\lbrace C_{Li},C_{Si}\right\rbrace + b
\end{equation}
By tuning the $a$ and $b$ coefficients, it is possible to see (Figure \ref{fig:mossbauer}) that this simple dependence yields the qualitative experimental trend (pink line). However, the definition given in equation \ref{eq:C} depends on the average of Li content of the alloy, while MB senses the local environment. We can seek a better agreement if we calculate the local concentration of Li and Si atoms by considering the nearest-neighbors of each Si atom for each of the amorphous alloys, as obtained from the averages of the simulations. If we replace into equation \ref{eq:mossbauer} the $C_{Li}$ and $C_{Si}$ values obtained from the simulations in this way, we get the cyan dots of Figure \ref{fig:mossbauer}, showing an improved agreement with experiment.

We can summarize the results obtained herein by stating that we have used a recently developed DFTB model to generate the configuration of amorphous Li-Si alloys of different Li contents, covering the usual experimental range. These configurations lead to pair distribution functions that are in excellent agreement with experimental results. When the same configurations are used to predict NMR and Mössbauer spectra at different alloy compositions, also a very good agreement is found with experiment. The present results encourage to use this new DFTB to predict other properties of Li-Si alloys, a promising system in the field of Li-ion battery materials.

\begin{acknowledgments}
This work used computational resources from CCAD-UNC, which is part of SNCAD-MinCyT, Argentina.
F.F. thanks his PhD fellowship from CONICET. We acknowledge financial support from CONICET (28720210100623CO, 28720210101190CO, 1220200101189CO, PUE/2017), the Agencia Nacional
de Promoción Científica y Tecnológica (FONCyT 2020-SERIEA-02139, 2020-SERIEA-03689) and SECyT of the Universidad Nacional de Córdoba.
\end{acknowledgments}

\bibliography{apssamp}


\setcounter{equation}{0}
\setcounter{figure}{0}
\setcounter{table}{0}
\makeatletter
\renewcommand{\theequation}{S\arabic{equation}}
\renewcommand{\thefigure}{S\arabic{figure}}

\begin{widetext}
    \begin{center}
        \textbf{\large{SUPPLEMENTAL MATERIAL}}
    \end{center}

\begin{figure*}[h!]
    \centering
    \includegraphics[width=\textwidth]{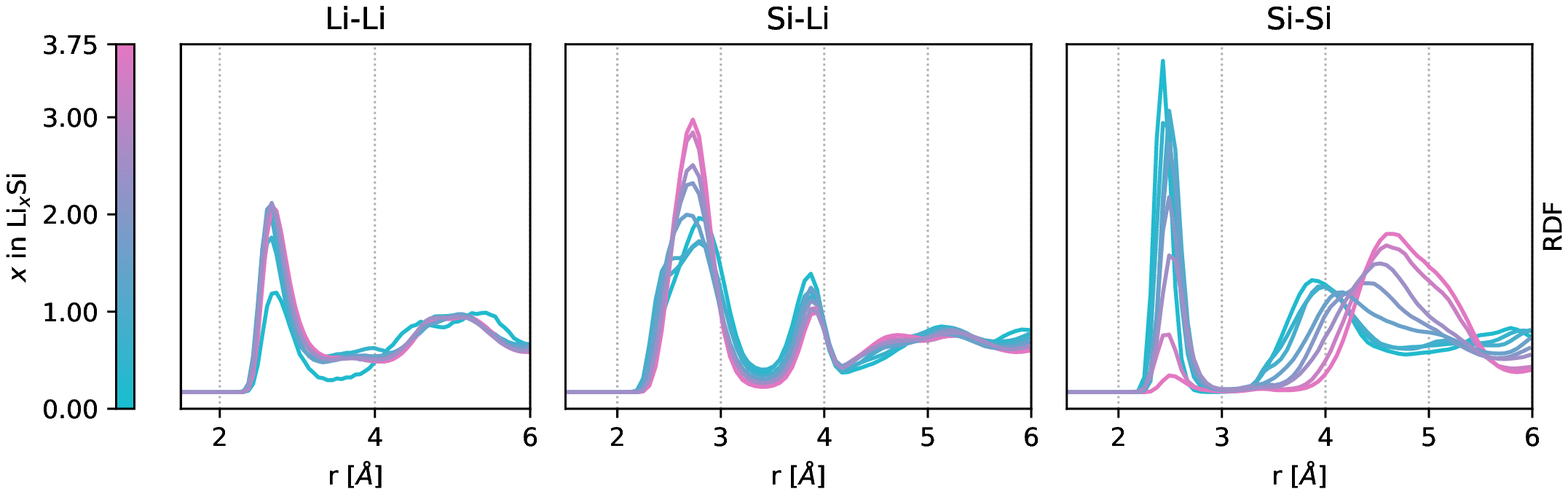}
    \caption{\label{fig:rdfs}Partial RDF for LiLi, SiLi and SiSi for all Li 
    concentrations considered. The colour of the curve changes from cyan (Si predominance) to pink (Li predominance).}
\end{figure*}

\begin{figure}[h!]
    \centering
    \includegraphics[width=0.8\columnwidth]{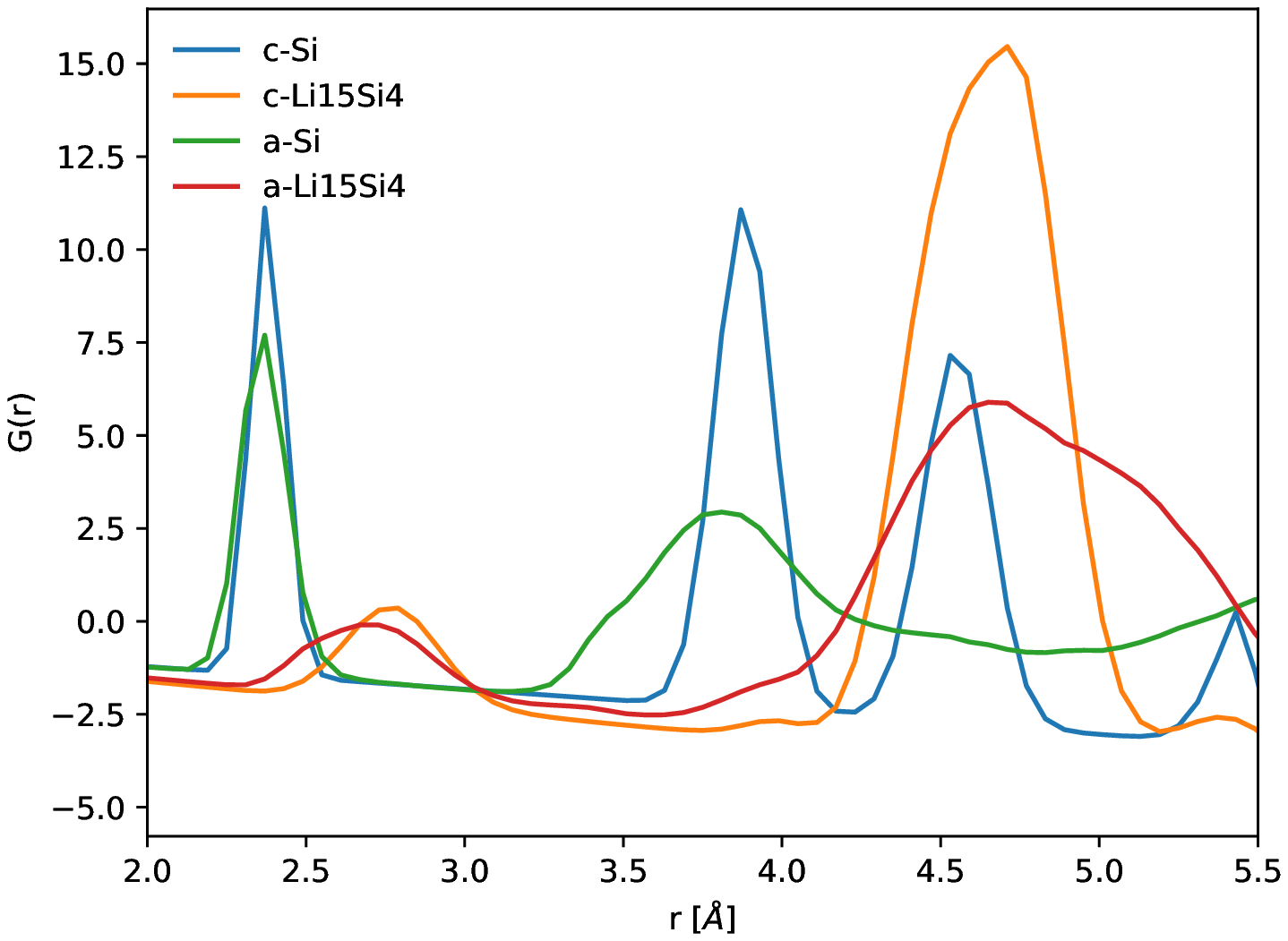}
    \caption{\label{fig:gofr-contributions} Pair distribution functions $G(r)$ of crystalline and amorphous structures used to compute the curves shown in Figure 1 of the manuscript.}
\end{figure}

\begin{table*}
  \caption{Weight factor of each contribution (c-Si, c-Li$_{15}$Si$_4$, a-Si and a-Li$_{15}$Si$_4$) to the pair distribution function G(r) of lithiated Si, shown in Figure 1 of the main text. The percentage represented by each weight is shown in parentheses. G(r) was calculated using equation \ref{eq:contributions}.}
  \centering
  \begin{tabular}{ccccc}
    \toprule
    Fitting & c-Si & c-Li$_{15}$Si$_4$ & a-Si & a-Li$_{15}$Si$_4$ \\
    \midrule
    crystalline & 0.03358 (30.15\%) & 0.077784 (69.85\%) & -- & -- \\
    DFTB amorphous & 0.0 (0.0\%) & 0.036422 (9.78\%) & 0.187971 (50.48\%) & 0.147955 (39.74\%) \\
    \bottomrule
  \end{tabular}
  \label{tab:contributions}
\end{table*}

The $G(r)$ of the lithiated Si structure has the contributions of each structure $S$ in $\lbrace$c-Si, c-Li$_{15}$Si$_4$, a-Si, a-Li$_{15}$Si$_4\rbrace$ multiplied by a factor $w_S$ that were adjusted to minimize the mean squared error and are summarized in Table \ref{tab:contributions},\begin{equation}\label{eq:contributions}
    G(r) = \sum_S w_S \cdot G_S(r),
\end{equation}
where each $G_S(r)$ is shown in Figure \ref{fig:gofr-contributions} and can be calculated using that
\begin{equation}
    G_S(r) = 4 \pi r \rho_{0,S} \left[g_S(r) - 1\right], 
\end{equation}
where $\rho_{0,S}$ is the atomic number density and $g_S(r)$ is the atomic radial distribution function of the structure $S$ and can be computed by considering the scattering factor $b_i$ of each atom $i$ in the following way~\cite{billinge2019}
\begin{equation}
    g_S(r) = \frac{1}{4 \pi \rho_0 r^2 N} \sum_i \sum_{j \neq i} \frac{b_i b_j}{\left\langle b \right\rangle^2} \delta (r - r_{ij}), 
\end{equation}
where $\langle b \rangle$ is the average scattering factor, $N$ the total number of atoms and $r_{ij}$ is the distance between atoms $i$ and $j$. The scattering factors for Li and Si atoms are $3$ and $14$, respectively. This gives us that the Si-Si RDF contributes in a 82\%, the Si-Li RDF in a 16\% and the Li-Li RDF in a 3\%.

\begin{figure}[h]
    \centering
    \includegraphics[width=0.8\columnwidth]{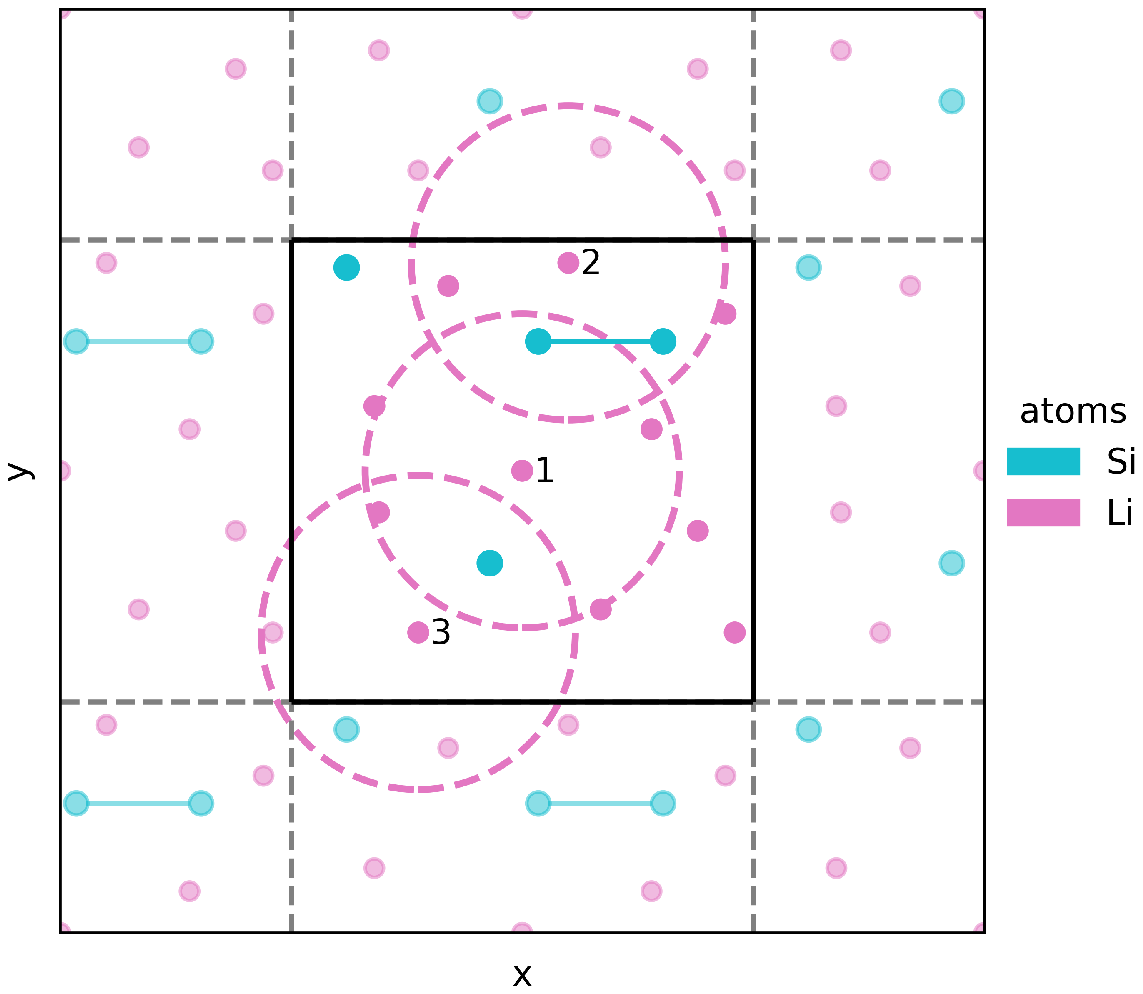}
    \caption{\label{fig:nmr-explanation} Diagram to explain the nearest-neighbor model that predicts the results of NMR experiments on LiSi systems.}
\end{figure}

In Figure \ref{fig:nmr-explanation} a simplified 2D unit cell with periodic boundary conditions is presented to explain the nearest-neighbor model that predicts the results of NMR chemical shift spectra for LiSi systems. Each Li atom contributes with a Voigt peak, as stated in the main text, where it centers depends on the type of Si atoms (bonded or isolated) that are in its first coordination shell. The cut-off radius for the first coordination shell can be seen in the Si-Si RDF in Figure \ref{fig:rdfs}, it is seen that the same is maintained for all concentrations at 3.4\AA. For example, for Li atom $1$ we have
$$
\delta_1 = \frac{18 \text{ppm} + 6 \text{ppm}}{2} = 12 \text{ppm},
$$
where the 18 ppm and 6 ppm come from the bonded and the isolated Si atom, respectively. In the same way, Li atom $2$ has a 
$$
\delta_2 = \frac{18 \text{ppm} + 18 \text{ppm}}{2} = 18 \text{ppm},
$$
and Li atom $3$ a
$$
\delta_3 = \frac{6 \text{ppm} + 6 \text{ppm}}{2} = 6 \text{ppm}.
$$
Then, the intensity of the spectra generated from these three atoms has is a sum of contributions
$$
I = V_1(12\text{ppm}) + V_2(18\text{ppm}) + V_3(6\text{ppm}) + ...,
$$
and so on as we consider the rest of the Li atoms in the structure.


\end{widetext}

\end{document}